# RFID Localisation for Internet of things Smart Homes: A Survey

Belal Alsinglawi[1], Mahmoud Elkhodr[1], Quang Vinh Nguyen[1,2] and Upul Gunawardana[1], Anthony Maeder[3] and Simeon Simoff[1,2]

[1]School of Computing, Mathematics and Engineering, Western Sydney University, Sydney, Australia
[2]MARCS Institute, Western Sydney University, Sydney, Australia
[3]School of Health Science, Flinders University, Adelaide, Australia

## ABSTRACT

*The Internet of Things (IoT) enables numerous business opportunities in fields as diverse as e-health, smart cities, smart homes, among many others. The IoT incorporates multiple long-range, short-range, and personal area wireless networks and technologies into the designs of IoT applications. Localisation in indoor positioning systems plays an important role in the IoT. Location Based IoT applications range from tracking objects and people in real-time, assets management, agriculture, assisted monitoring technologies for healthcare, and smart homes, to name a few. Radio Frequency based systems for indoor positioning such as Radio Frequency Identification (RFID) is a key enabler technology for the IoT due to its cost-effective, high readability rates, automatic identification and, importantly, its energy efficiency characteristic. This paper reviews the state-of-the-art RFID technologies in IoT Smart Homes applications. It presents several comparable studies of RFID based projects in smart homes and discusses the applications, techniques, algorithms, and challenges of adopting RFID technologies in IoT smart home systems.*

## 1. INTRODUCTION

The Internet of Things (IoT) foresees the interconnection of billions of things by extending the interactions between humans and applications to a new dimension of machine-to-machine communications. Rather than always interacting with the users, things will be interacting with each other autonomously by performing actions on behalf of the users [1]. The IoT provides the user with numerous services and capabilities. The obvious ones are the ability to control and monitor the physical environment remotely over the communication networks. Typical examples are the ability of closing a door or receiving smoke alert notifications remotely over the Internet. However, the revolution in technology actually occurs when things and group of things are connected together. The interconnection of things allows not only things to communicate with each other, but also offers the opportunities of building intelligence and pervasiveness into the IoT. The interconnected network of things, along with backend systems involved in a number of collaboration activities with the users and other things, in tandem with cloud computing systems,

Big Data, web services, and Location Based Services, will transform not only communications on the Internet but also societies [1].

The smart home is an area in which the IoT promises to reshape. The IoT enables everyday household objects, electronics and smart appliances to communicate with one another either locally or via the Internet [2]. Most of the existent smart home systems use devices at the lower end of the envisioned capabilities. Typically, these devices are capable of storing data, responding to user commands from smartphones, tablets and computers, and sending alerts over Bluetooth or Wi-Fi. They generally operate in a standalone manner. The IoT brings a new type of home management, integration of devices, surveillance, intelligence and more importantly connectivity to these devices. Intelligence may be contained completely within a device, combined with platform intelligence in the cloud, or reside almost completely within a platform to which the device connects to perform some functions. Therefore, smart home devices incorporate the capabilities inherent in the IoT and provide enhanced benefits. Smart home devices can be static objects, such as smart plugs or lights that simply report their properties. They can also be sensors that measure the physical conditions of an object or its status, actuators that perform operations (opening doors, turning on or off appliances), or devices that combine both of these services. Significantly, the IoT system will enable these devices to be queried or controlled by other platforms, controllers, or IoT applications that coordinate multiple objects without the interference of the human user. In addition, the data collected from smart home devices can be integrated with external data collected from other IoT systems, e.g. a healthcare system, which create value-added services. Therefore, the user benefits from added intelligence, modelling, and weaving of information, which enable the smart home system to make better decisions on behalf of a user, or to provide personalized and optimized services. This cannot be achieved without the integration of context-aware technologies in the IoT. Mainly, contextual data in the IoT is used to provide tailored services, increase the quality/precision of information, discovery of nearby services and making implicit users' interactions [3]. Thus, localisation is regarded as a key enabler for this technology.

Localisation in indoor environments has gained popularity in the domain of ubiquitous computing in the last few years and will continue to play an important role in the IoT. Indoor Positioning Systems are systems that use wireless communication networks (short-range to long-range) [4]. Several technologies in indoor environments have been adapted to different applications such as asset management [5], IoT healthcare, security, warehouse and people tracking. Technologies such as RFID, Bluetooth, Wi-Fi, among many others are commonly used in IoT Smart home applications. RFID is the one technology in particular that promises to revolute many industries due its low-cost and low-power characteristics. While the adoption of the RFID is progressing fast, many challenges are still need to be addressed.

To this end, this paper reviews the state-of-the-art technologies in IoT Smart Homes and comprehensively evaluates and studies existing RFID based projects. Section 2 discuss developments of smart home systems and RFID technology. It further discusses the challenges pertaining to the adoption of RFID based localisation solution in IoT smart home applications. Section 3 reviews several RFID solutions and localisations techniques. It also provides several analytical and comparative studies of traditional RFID based solutions for Smart Homes. Section 4 provides the conclusions for this work.

International Journal of Computer Networks & Communications (IJCNC) Vol.9, No.1, January 2017

## 2. SMART HOMES SYSTEMS AND RFID

The principle of localisation systems in Smart Homes (SH) applications depends on sensing the activity performed by individuals and locating the positions of movable entities. Localisation technologies can be categorised to as follows:

- Radio Frequency based technologies.
- Optical sensors.
- Sound waves sensors.
- Electromagnetic field sensors [6]. These technologies are commonly used in SH settings for subjects tracking and objects localisation. RF based systems have gained significant popularity in various smart environments applications.

Radio Frequency Identification (RFID) is a desirable technology for localisation in the IoT. RFID Systems consist of Antenna's connected to RFID readers. These antennas send back the captured information from sensed tags to the reader for further location processing. RFID tags can be Active tags (battery powered), semi-active and Passive tags (without a built-in battery). Tracking of RFID systems can be active or passive. There are various RFID tracking applications such in-hospital patient tracking, asset tracking, supply chain, security, medical and healthcare assets tracking.

Recently, RFID technologies have been widely deployed in modern logistics and inventory systems for efficient monitoring and identification [7]. This is because RFID technology is considered low-cost, usable, and provide a reliable form of automatic identification, which makes it a cost effective technology to use for localisation in indoor environments. Furthermore, RFID has favourable characteristics such as contactless communication, security and a high data rate and non-line-of-sight readability [1]. However, issues such as variations in the RFID reader's data, RFID tags performance, interferences among other tags related problems pose some serious challenges to the adoption of RFID localisation technologies in IoT smart homes. RFID works by sending and receiving the unique identity of persons and objects wirelessly by using radio waves. An RFID system consists of readers, tags and a data collection module as seen in Figure 1. The readers can be static or mobile. There are two methods for tracking. In the first method, the reader can be installed in a static location inside the household (such as a wall, a table or a kitchen) to sense the movement of RFID tags. The reader then searches for the tags which are either attached to objects or carried by the person. In the second method, the portable reader detects the static tags in certain positions while the RFID reader can be carried by individuals [8, 9].

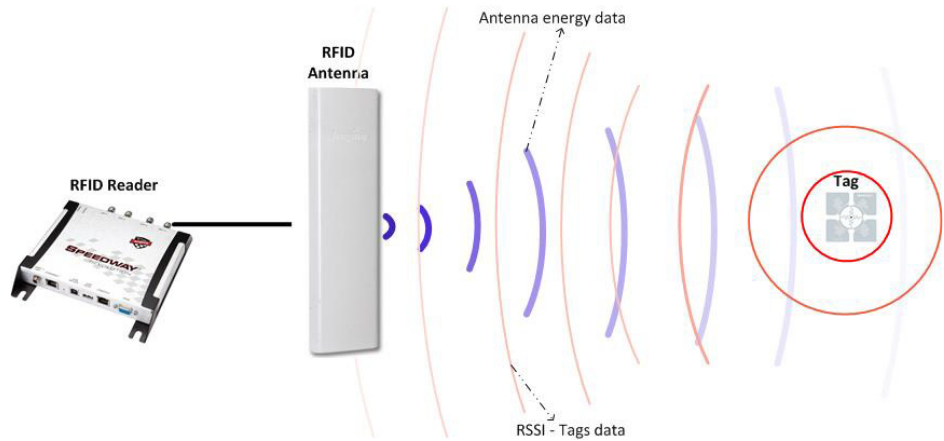

Figure 1. Passive RFID Localisation System Architecture

There are three main categories of RFID tags including active, passive and semi. Active RFID tags have an internal battery to power itself continuously. They have the greatest range of all three types [9] However, active tags lifetimes are limited and they rely on how much energy is stored in the internal batteries. The maintenance and intrusiveness levels for active tags are much higher than for their passive counterparts. The price of active tags and the maintenance cost are relatively higher than other types. Semi RFID tags also have an internal battery to power the internal circuity [9].

Passive RFID tags have no internal battery. They are smaller in size and are much cheaper than active or semi-active tags. Notably, passive RFID tags are powered by the radio waves that are emitted by the antennas so they do not have an internal source of power. The tags are usually applied on objects in smart homes, such as cups, kettle or furniture [10]. They support elderly citizens who live independently in their residential homes and need less expensive service than traditional nursing homes or hospitals.

## 2.1 RFID Localization Techniques in Smart Homes

Several RFID indoor localisation methods have been proposed in the literature. There are three main detection techniques and position estimations for RFID technology including Triangulations (distance estimation), Scene Analysis, and Proximity.

### 2.1.1 Distance Estimation

This technique relies on the geometrical properties of triangles to determine the target locations as shown in Figure 2. In Distance Estimation, the lateration method estimates the position of an object, where RFID tags are attached, by measuring its distance from multiple reference points (either using RFID objects or RFID antenna). This technique is usually called the range measurement technique [11].

On the other hand, a time-based method, such as time of arrival (TOA) and time difference of arrival (TDOA), are techniques which measure the positions of RFID tags (or objects) based on distance measurements [11]. The received signal strength (RSS) is based on the received signal phase method and the phase of arrival based techniques such as the phase difference of arrival (PDOA) [12]. Other technique known as Trilateration & Multilateration, both require several spatiality reference points (using an RFID antenna) to perform a position estimation. These techniques measure the distance value of each reference point obtained by converting RSSI to a computed distance measurement.

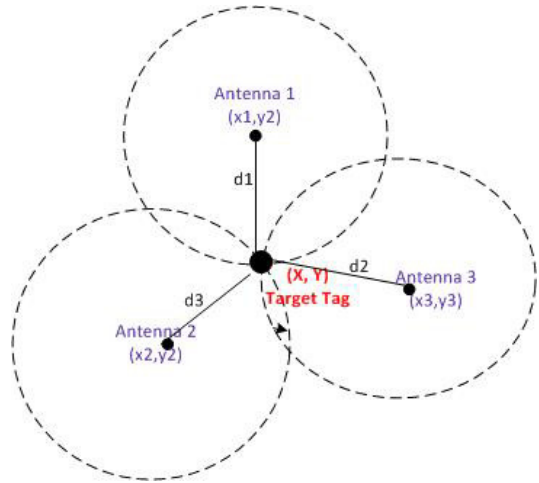

Figure 2- Location estimation using Trilateration Algorithm

Many researchers have investigated RFID localisations using the triangulations on various models for indoor positioning such as TOA [13], TDOA [14], PDOA [15] and RSS [16] [17]. Localisation, along with RFID Lateration techniques have been used for various indoor positioning applications. However, they suffer from some drawbacks such as multipath (TOA,TDOA) and non-LOS ( TOA, TDOA and PDOA). The Angulation technique e.g. Angle of Arrival (AOA) is defined as the angle between the propagation directions of the incident waves and references, which is called orientation. The orientation is defined as the fixed direction against which AOA is measured [63]. In this approach, the location is determined in 2-D by calculating the intersection between two beacons emitted by the tags, and the their positions as measuring elements [18]. AOA requires two beacons to improve the accuracy and more than three angles for triangulation. However, AOA is affected by multipath, NOLS propagation and wall reflection, which causes errors for indoor location estimation [12].

**2.1.2 Scene Analysis**

The Scene analysis method collects the features (fingerprints) of a scene and then estimates the location of the tagged objects by matching the online measurements with the closest deductive location fingerprints [11]. One of the most common approaches is RSS location based fingerprinting. There are two stages in the location fingerprint: the offline stage and the online stage (run-time stage). A site survey is performed in certain environments during the offline stage. The locations of coordinate values or labels and signal strength are determined by collected near measuring units. In the online stage, the current detected signal strength and the information gathered to discover the new estimated location. However, errors can happen in received signals whose strength can be influenced by the reflection, diffraction and scattering that occurs in indoor environments [11]. Fingerprinting-based positioning methods typically consist of five

International Journal of Computer Networks & Communications (IJCNC) Vol.9, No.1, January 2017

pattern recognition techniques including probabilistic, k-nearest-neighbour (kNN), neural networks, support vector machine (SVM), and smallest M-vertex polygon (SMP) [11].

### 2.1.3. Proximity

In this method, the location depends on the symbolic relative location that comes from an intense grid of antennas. When a mobile target enters the single antenna's radio signal range, the antenna will consider the target as a collocate object on its entire coverage. If more than one antenna detects the same target, it will be collected by the antenna that receives the strongest signal. The cell of origin (COO) defines the position of the mobile target and the positon is within limited coverage. The localisation method is simple and it does not require heavy implementation. However, the accuracy relies on the density of the antennas and the signal range. This also means that the approximate position of the tagged object is used at a given time.

### 2.2. Issues & Challenges

By integrating the smart home system with the IoT, IoT devices, such as RFID tags will be capable of communicating with each other. These devices will also be capable of obtaining information about other devices; or they will be capable of controlling the physical environment based on the user's preferences. They can use information external to their specific environments to enhance the operation of the IoT smart home system to the user's benefits. Ultimately, smart homes will evolve from traditional smart systems to sophisticated environments that provide assisted living, support in-house remote health monitoring services and elderly monitoring [1]. They can be used to provide solutions for monitoring patients at home. These systems can deliver a higher quality of care and reduce the cost on patients and governments without affecting the quality of the healthcare services provided. For example, the use of a remote monitoring system allows biomedical signals of a patient to be measured ubiquitously during his or her daily activities. Such a system allows the collection of medical data and signals related to patients' bodies, such as their heart rates, current activity and location, remotely via the Internet.

An IoT based remote monitoring system is capable of detecting any changes in the person's body conditions, and monitoring their vital medical signs. The availability of the collected data by this system on the Internet, and the ability to access this data in real-time by various other systems and entities such as healthcare providers and medical centres, open the door to numerous opportunities. For instance, an alert system can be designed based on analysing the data received by the remote monitoring systems. In the case of a medical emergency, the system can be configured to alert the healthcare professionals, emergency services, relatives and others concerned parties. Also, the system can provide insight into the health condition of a monitored person so the necessary help can be provided as early as possible, and thus, saving patients' lives. However, for Smart Home systems, in general, and remote health monitoring systems, in particular, to process automatically information about peoples' locations such as their movements around the house, these systems require to track the locations of people accurately. However, energy efficient indoor positioning solutions remain an open research challenge.

### 2.2.1 Accuracy Challenges

Location accuracy is one of the biggest challenges that Smart Homes settings are facing. There are multiple factors that affect precision of the sensed location including:

- The method used for determining locations of the subjects in indoor environment.

86

- The capability of the device and the technology in use.
- The size of the deployment area
- The distance between targeted objects and sensing devices (e.g. sensors and readers) and physical obstructions.
- The orientation of tagged objects from sensors also plays a significant role in obtaining a better location measurement.

### 2.2.2 Low-cost Challenges

Low-cost and low-power are vital features in the IoT. The cost of a solution includes the costs of infrastructures, positioning devices and system components, installations and maintenance, and the costs associated with testing and calibrations. While, WLAN systems have a relatively good accuracy that can be used as addition to existing localization devices, RFID systems are more desirable as they use cheaper devices compared to those used by WLAN [11].

### 2.2.3 Complexity Challenges

Complexity of the solution is an important factor when designing smart homes. Adding more infrastructure components to the existing solution will result in more complex system. If a system requires more time to locate the subjects during the localization and with a complex algorithm, the system is likely to be effective. Long term maintenance adds also to the complexity of maintaining this system especially if the system requires frequent maintenance.

### 2.2.4. RFID Localization Systems Challenges

The main challenges of RFID Location tracking systems and technologies are the high variation of principles and functionalities in localising objects and moving subjects in indoor environments. Researchers have worked on finding optimal indoor localisation solutions that worked on numerous indoor positioning platforms. Nonetheless, there is no fully optimal solution using RFID platform technology.

### 2.2.4.1 Variation in Tags and Reader readings - Behavioural Challenges

One of the most common problems that is facing passive RFID tags being used in tracking systems is the fact that passive RFID tags fluctuate in their RSSI readings, even if the tags and readers are static (in a fixed position) and no objects or subjects are crossing one another. Furthermore, tags that are working in the same conditions may be different in RSSI. These behavioural variations could be caused by manufactural defects or even differences within chips, integrated circuits and noise [8]. More approaches and methods need to be undertaken to be able to calculate the RSSI changes and RFID tags abnormal behaviour.

Another common issue in RFID localisation systems is RFID Readers behavioural variations in RSSI signals. This happens when the readers are not able to fully query the tags within their reading range [19]. This could be addressed by finding a mechanism that is able to increase the power level and also find ways to optimise the distance between the tags and readers within acceptable reading values and without significant changes in RSSI readings.

**2.2.4.2 Interference Challenges**

Interference is a common issue in RFID localisation. It is caused by environmental interference factors like radio noise and collision caused by metal and liquids which is an impermeable for the signal to pass through. Internal factors related to RFID such as tags, readers can also create interference. The interference causes RF propagation and eventually lead to error in localisation [20]. The interference problem can affect both active and passive tags in localisation. However, in active mode tracking, the localisation errors are less than in passive tracking because the active RFID readers emit less energy to detect the tags. In passive tracking, RFID readers require more energy to localise the passive RFID tags that do not have any source of energy and rely on the RFID readers' emissions. UHF RFID interference can be divided into three types including tag interference, multiple readers to tag interference and reader to reader interference [21]. Research has proposed to reduce localisation errors caused by interference such as those in [8] [22]. Unfortunately, further investigation is required to produce better and more scalable results.

**2.2.4.3 Other Tags Related Challenges**

Tags orientation is very important for detecting tag's locations via reader communication. Tags can be attached vertically, horizontally or at an angle on the sides of objects to obtain better detection. Parallel orientation usually reduces the detection chances compared to the previous setup due to one side of directivity in parallel orientation.

Sensitivity of the tags is another issue in RFID localisation applications. It defines the minimum power required to activate or read the tags. Tags with lower sensitivity cause more locations errors while tags with higher manufactural sensitivity provide better location detection [8].

Tag spatiality affects the localisation errors. For example, frequent replacement of the tags at random locations will lead to lower accuracy. Tags placement at certain locations will give better results during the interactions between readers and tags.

**2.2.5. Traditional RFIDs Solutions**

Table 1 compares various common RFID localisation algorithms in Smart Homes. Each localisation technique has one or several localisation methods. In general, all techniques cater for two dimensional localisations whilst only few localisation techniques are capable to track in both 2 dimensions and three dimensions. Furthermore, some of these techniques provide better location precision. Therefore, choosing and adapting a suitable localisation method to a given scenario or application relies on the system requirements in indoor environment and the localisation application area, such as human localisation for wellbeing, tracking, medical equipment, supply chin, goods tracking and many indoor localisation applications. Table 2 presents the most common technique in Scene analysis localisation.

Table 1. A comparison between the common RFID localization methods.

| RFID localisation Technique | | Method | Dimension | Advantages (A) / Disadvantages (D) | Reference |
|---|---|---|---|---|---|
| Distance Estimation | *Lateration, Triangulations, Time based, phase based and Tag Range based Techniques* | TOA | 2D | A : High precision localisation  D: direct TOA suffers from synchronization and time-stamp multipath effect. | Shen et al [13] |
| | | TDOA | 2D | A: Accurate for Real time locating (RTLS)  D: NLOS Multipath | Kim et al [14] |
| | | POA/PDOA (RSP) | 2D | D: Multipath propagation Rely on LOS | Povalac [15] |
| | | RSS | 2D,3D | A: -Cost effective method of location estimation  -Better estimate of the distance.  D: uncertainty location related issues | Chawla et al [16] |
| | Trilateration and Multiliteration | | 2D, 3D | A: high level of accuracy  D: - Require to use at least reference points to perform distances calculation  - measurement errors, | Bouchard et al [23]  Alsinglawi et al [24] |
| | *Angulation* | AOA | 2D,3D | A: no synchronization required  D: multipath reflections | Azzouzi et al [25] |
| Scene Analysis (fingerprint) | | **(Refer to Table 2) | 2D,3D | | |
| Proximity | | Reference Points (Well-Known position) | 2D | A: offer proximate position information  D: cannot give absolute (relative) position | Song [26] |

Table 2. Algorithms in RFID Scene Analysis (Fingerprint).

| Scene Analysis Algorithms | Description | Author |
|---|---|---|
| Probabilistic Approach | Based on Bayesian network [86] to estimate target (tags) location. | Seo et al [27] |
| k-nearest-neighbor (kNN) | Radio mapping based in online RSS. | Ni et al [28] |
| Neural Networks method | It uses offline RSS and a-like location coordinates as an input for the target training purpose. | Moreno-Cano et al [29] |
| Support Vector Machine (SVM) | It uses statistical analysis and machine learning to perform the classification and regression. | Yamano [30] |

## 3. LOCALIZATION SYSTEMS IN SMART HOMES

Localisation of objects and persons is still a challenging issue in SH. Researchers trying to find inexpensive solutions for tracking objects in the indoor environment using RFID or different locating systems. Some of these works are hybrid approaches that provide better localisation yet add cost and complexity to the systems. Unfortunately, indoor positioning in smart homes is still a difficult task, especially with an affordable solution.

Many research projects have been proposed in the quest of finding a solution in smart homes. Many factors were considered in those projects, such as accuracy, precision, cost, complexity and adaptability of the system as well as robustness and scalability in indoor positioning. RADAR [11] by Microsoft is the first RF project that was applied in an indoor localization environment. RADAR was based on Wi-Fi signal strength or fingerprinting localization method. The system achieved an accuracy between 2m-5m with 90% precision within 5.9m [11]. It uses WLAN networks infrastructure which is also easily installed. Nevertheless, the localization devices have limited energy levels and received signal strength (RSS) as well as privacy concerns [11].

Cricket [18], is the first noble indoor localization system that combines ultrasound with RF technologies that could be used in various indoor applications such as medical, healthcare and human tracking. The system can achieve high accuracy between 1-3cm for long range tracking with reasonably low costs. However, this system uses battery powered tags that are not an ideal solution for long running times, whereby the system suffers from an energy consumption issue. Active Badge [31] is a novel location system proposed by researchers at MIT University. It aims to detect the location of staff members inside the university facilities and provides information about their movements. Active badge is a cost effective solution based on Infra-red (IR) sensors which uses a small wearable device to transmit IR every 15 seconds to the sensors through an optical path. Interestingly, Hodges et al. [32] proposed a system that uses ultrasonic sensors to determine the location in 3D. The system achieves up to 95% efficiency when reading at 3cm. However, the implementing of the sensors is relatively costly.

### 3.1. RFID Localization Projects in Smart Homes

Early research conducted was focused on RFID technology for indoor tracking but RFID has also been applied to many areas such as the industrial, medical, automobile and agriculture fields. RFID technology has been scientifically proven in applications thanks to its advantage in accuracy, cost, efficiency, adaptability, scalability, robustness and low complexity. A comprehensive list of RFID solutions for smart homes (SH) is discussed in Table 4.

Ni et al [28] introduced the concept of localisation using tag references in their LANDMARC system. The LANDMARC system uses active tags that are located in fixed positions. The system measures the distances between readers and tags using a multi-level power method. There are eight levels where level 1 is the shortest range, and 8 is the longest. The system relies on the signal received to estimate and detect the position of the tags. LANDMARC obtains an accuracy of 1m (50% error distance) with less than 2m in maximum error distance. Work by Zhao et al [33] used the principles of the above LANDMARC system to implement the Virtual Reference Elimination (VIRE) system that locates reference tags in a virtual reference tag and enhances the performance to avoid interference as well as multipath. The authors reported the least error estimation 0.47m within the average of error estimation (0.29m) for non-boundary tags [33] when compared to LANDMARC. The works in [34] and [35] also usedthe LANDMARC's concept to enhance the localisation. Jin et al [34], which improved the overall localisation performance of the former and achieved an accuracy of 72cm using fewer tags around the targets. FLEXER [35] utilised a simulated solution to increase the accuracy where their solutions attained 70cm (or 80%) using applied region mode [35].

Zhang et al [36] implemented an RFID diversity elimination algorithm called RFID DeffFree Loc to reduce the mean locating error. In their simulated work, the system obtained accuracy of 10cm in free noise, while 19cm was achieved in noisy environments and the accuracy was improved in noisy environments compared to LANDMARC solutions. The system obtained accuracy of 10cm in free noise, while it achieved an accuracy of 19cm in noisy environments. In contrast, Work by Hahnel et al [37] was one of the early projects that considered indoor localisation and mapping using passive RFID tags based on a probabilistic measurement model. It applied two antenna readers installed on a mobile robot to detect the static passive RFID tags that were attached to the walls of the tested environment.

Tesoriero et al [38] expressed the idea of turning the area (floor surface) into a grid. A RFID reader attached to a mobile robot sensed the passive RFID tags that were also attached in small spaces inside the grid floor. RFID tags were linked to a particular position on a virtual map. The system achieved an accuracy of 0.9m. However, the readers have to be carried during localisation.

Hekimian-Williams et al [39] proposed a project that achieved very high and precise localisation results in millimetres using Phase Difference between two readers. The system used a simple approach that consisted of 2 readers for locating 1 active RFID tag. The system however always used battery powered tags which require high maintenance and had associated high costs. Systems such as in [17, 20, 40] also provided high accuracy in their solutions Vorst [40] applied a Particle Filter (PF) based on a pre-probabilistic approach (self-localisation) to achieve better accuracy while Joho [17] used an antenna orientation and RSSI model. Similarly, Chawla [20] developed several new linear, binary and parallel search localisation algorithms to enhance the overall accuracy and achieved very good results compared to the previous works with up to 18cm accurate localisation. However, these systems use many tags and more readers which in turn added costs as well as added to the complexity of the system which make them less suitable for indoor environments.

Hybrid methods that combined UWB with RFID were also proposed. Semi-active tags were introduced by D'Errico R. et al [41] where the system used UWB antenna and UHF technology to sense RFID Semi-active tags based on backscattered signals received modelling. The system detected two types of tags 1) dynamic tags based on extended Kalman filters (EKFs) algorithm and 2) static tags using a least squares (LS) algorithm. This method achieved relatively good accuracy (20 cm) with less than 0.53m (75%).

Xiong et al [42] combined WSN and RFID devices as a hybrid approach to achieve desirable results. The authors applied hybrid cooperative positioning algorithms that extended Kalman filter (EKF) with numerous measurement modules [42]. Their purpose was to find a reliable solution to indoor positioning that was compatible with existing infrastructure from different IPS technology. The method was tested in simulated and experimental environments and achieved considerable levels of accuracy.

Fortin-Simard et al [43] proposed a method that adapted a new, enhanced trilateration approach using RSSI. They applied various filtering algorithms to reduce the localisation errors that were caused by the interference of environmental surroundings (e.g. metal and walls). The work also reduced the problems generated by the nature of the passive tags and RFID readers during localisation processing. It obtained a high accuracy of 14cm overall. The system was also implemented to support daily life activity recognition. Similarly, the work by Bouchard et al [23] was presented to enhance Fortin-Simard's trilateration model and algorithms [43]. The authors improved the fuzzy localisation using the mean of interface engine and linguistic variables such as likeliness, distance and object detection. Alsinglawi et al [24], proposed localisation framework using trilateration algorithm to assist with elderly localisation in Smart homes healthcare settings. The system reported promising accuracy results using only 3 antennas and one target passive tag. The system reported accuracy levels above %90 at the centre area of the localsaition platform with an average 16.5cm.

Sunhong et al [44] used RFID readers attached to a robot to detect fixed location tags on the floor during robot movement. The work aimed to provide assistance and localisation movement of elders and individuals with disabilities who used motorized wheelchairs. The researchers presented an algorithm to read the speed of the robots movements (or a portable chair), where the accuracy depended on the reading speed against the tag locations. This obtained a promising accuracy of 10cm in comparison to previous similar approaches.

Jachimczyk et al [45] utilised a 3D RFID localisation method using hybrid algorithms in scene analysis and neural network. The system was tested in three different test cases including active readers, different scenario and cost effective. It performed in both simulated and in real environments to find the optimal configuration for RFID readers. The results were obtained according to a number of active readers in different scenarios. The scenarios required a certain number of RFID readers to be allocated in each test where it performed according to a certain number of readers (from 1 to 8 readers). The best condition was achieved when using four readers or 8 readers and the averages of the accuracy were 11cm and 7cm respectively as well as 49cm and 50cm respectively for standard deviation uncertainty.

Athalye et al [46] proposed a solution for indoor localisation by using new semi-active tags called senstags with dual detection ability. Senstags first detect and decode backscatter signals from RFID tags (within proximity range) and then communicates with the reader using backscatter modulation as a regular tag. Although this technique achieved good accuracy, it requires a long battery life and high system maintenance. Yang et al [47] introduced some principles for tag distribution localisation and grid patterns. The system defined SRE as the ratio of the number of successful tag readings. This method has been successfully applied in detecting multiple RFID passive tags. Bolic et al [48] presented an approach called Sense-a-Tags (STs) by applying the proximity technique to enhance the passive RFID tags functionality in tracking people and their interaction with objects in real time. The authors tested their STs system in two experiments using a particular number of RFID tags with various tag orientations. The system achieved 32 cm and 48 cm detection accuracy respectively.

Table 3 compares the common RFID and recent localisation approaches for indoor Smart Homes. The experimental results represented by (E), Simulation results (S), both: simulation and experimental (S&E). Target- location (H) for human tracking and (T), for tracking only tags.

Table 3. Comparing RFID based solutions for Smart Homes

| Solution | Application | Accuracy | Technique | Readers/tags | Benefits |
| | | Efficiency | Tracking | Area m/m²/m³ | Drawbacks |
|---|---|---|---|---|---|
| **LANDMARC [28]**, 2004[E] | Location *awareness* | ≤ 2m | References tags | 9/ 64 | - Cost effective solution.<br>- Less infrastructure required during deployment<br>-Minimises the localization error caused by environmental interference (more precise) |
| | | 1m (50%)<br>5.9m (90%) | Active tags | N/A | Complexity and flexibility such as:<br>- Long latency.<br>- Different tag behaviour during detection (different reading values) |
| **Jin et al,2006[34]**[E] | Location awareness | 72cm | Reference tags | 4/20 | New mechanism (based on previous work by LANDMARK) to reduce the computational load caused by tags (reduced number of neighbour tags). |
| | | 83cm (per 10 tags in 2m "average tolerance ") | Active tags | N/A | - System used changes active tags (high cost and battery requirement)<br>- Complexity and maintenance issues. |
| **FLEXER,2006 [35]**[S] | Location awareness (indoor localization) | 40cm-1m | Reference tags | 4/64 | - Flexible localisation method (localize region mode and coordinates)<br>- Reduces computational load and enhances the localisation speed |
| | | 70cm (80%) Region mode | Active tags | 49² | System used Active tags (high cost, battery requirement)<br>System complexity implementation |
| **VIRE[33]**, 2007[E] | indoor localization) | 1.5m | Reference tags | 4/16 | - Cost effective solution |
| | | 0.5m | Active tags | N/A | - Lack of the solution in large scale<br>- Complexity and maintenance issues (battery requirement) |
| **Zhang et al , [36]**[S] **2009** | location awareness | 10cm (1m space between tags | Reference tags | 4/49 | Reducing the diversities of tags in home environment and the mean locating error. |
| | | 19cm (2.4m space between tags) | Active tags | 100m | - High cost and system relies on active tag battery requirement).<br>- High complexity and needs maintenance. |
| **Hekimian-Williams et al[39]**[T],2010 | location awareness | Millimetres accuracy | Phase Difference | 2/1 | Very precise and highly accurate approach (accuracy in millimetres) |
| | | Very precise | Active tags | 18m | - Not applicable for tag localization for Passive tags.<br>- High cost (active tags are expensive and need *a* battery)<br>- High complexity and that needs maintenance.<br>- System suffers from intrusiveness that resulted from multipath. |
| **Hahnel et al[37]**[T], 2004 | indoor positioning (robot localization) | ≤ 2m | References tags | 2/100 | Map learning approach using Mobile Robot |
| | | 1m-1.4m | Passive tags | 28² | - Required several RFID tags |

International Journal of Computer Networks & Communications (IJCNC) Vol.9, No.1, January 2017

| | | | | | |
|---|---|---|---|---|---|
| | | | | | (high cost and high complexity)<br>- Line of sight issues (Laser range scan) |
| **Vorst et al [40], 2008** | indoor localization (mobile robot) | 20cm-26cm | Reference tags | 4/374 | - Proposed probabilistic fingerprint technique (in particle filter) for considerable accuracy. |
| | | ≈ 0.25cm 0.32cm(90%) | Passive Tags | 125m | - High cost with many tags and readers.<br>- High complexity due to the large number of tags and readers |
| **Joho et al [17], 2009** | Indoor localization (mapping) | 27cm-29cm | References tags | 1/350 | Probabilistic sensor model (Sensor calibration) based on RSSI to improve the accuracy of the system |
| | | ≈ 35cm | Passive Tags | N/A | High cost as adding more tags will add extra costs to the system |
| **Tesoriero et al [38], 2009** | Indoor Tracking (autonomous entities) | ≈ 0.9 m | Sense Analysis | 1/19 | - Locating objects based on entities (inside grids).<br>- Virtual mapping |
| | | Error = 0 ( 50% speed against 19 tags<br>Error = 10% (75% / 18tags)<br>Error = 20% (100% / 14 tags) | Passive tags | $43^2$ | - High cost as it requires many tags for more efficient and accurate localization.<br>- Usability issues.<br>- High complexity, every object (even smaller, cups, kettle, etc.) need to be attached to readers for localization. |
| **Sunhong et al [44], 2010** | Indoor Tracking (robot location) | ≈ 10cm | References tags | 1/198 | A method to reduce number of used tags and sensors. |
| | | N/A | Passive tags | 26m | Usability issues, limited localization application (not suitable for real time for non-disabled elderly individuals) |
| **Chawla et al[20],2011** | Indoor localisation (object localisation) | 0.18cm | References tags | 1/132 | Several algorithms to achieve higher accuracy and efficient solution |
| | | 0.35cm (overall average) | Passive tags | 8m | - Need to deploy a large number of tags for higher accuracy<br>- High complexity and installation issues |
| **D'Errico, R., et al. (2012)[41]**[S&E] | Indoor localisation (Real time tracking) | 20cm | TOA | 4/Many | Minimise energy consumption (battery) by enabling semi active tags with UWB antenna and improved synchronization. |
| | | 0.37m-0.53m (75%) | Hybrid (UWB-Semi-active tags) | N/A | - High cost (adding more tags and readers will increase the cost of the whole system)<br>- Line-of-sight and multipath problems<br>- Interferences<br>- High complexity and maintenance issues<br>- Usability issues |
| **Fortin-Simard, D., et al. (2012)[43]** | Indoor localisation (Real-time tracking) | ≈ 14cm | Trilateration/RSSI | 4/4 | New trilateration positioning model with various existing filters and fuzzy logic to achieve accuracy and system efficiency. |
| | | ≈ 32.5cm ( higher efficiency) | Passive Tags | $6m^2$ | - Results obtained in limited coverage area (no actual test for various objects in smart homes e.g. furniture, different sized and shapes)<br>- Limited to positioning simple |
| | | | | | objects and does not cover multiple objects. |
| **Yang, Wu et al. 2013[47]** | Location awareness | 10cm | References tags | 4/96 | High accuracy based on tag distribution (grid approach) |
| | | 10cm± 2.56 cm | Passive tags | N/A | The results are varied upon different localisation algorithms and RFID tags. |
| **Athalye, Savic et al. 2013[46]** | Location awareness | 30cm | References tags | 1/12 | New Sense tags which have a dual ability to locate objects |
| | | ≤ 40cm CDF Method | Semi-Active | 6m$^2$ | Battery life issues caused by comparator that runs whole power circuit. |
| **Xiong, Song et al. 2013 [42]$^{E\&S}$** | Indoor Tracking (people / objects) | 1.6m | RSSI | 4/N/A | - Cost effective approach (combined WSN with RFID devices)<br>- Robust IPS solution (effective solution in harsh environment) |
| | | 1.8m (hcEKF algorithm ) | Hybrid RFID Passive Tags/WSN | 300$^2$ | System was not tested in *a* large scale experimental space. |
| **Bouchard, Fortin-Simard et al. 2014[23]** | Indoor tracking (people) / activity of daily living (ADL) detection | ≈ 16cm | Trilateration/RSSI | 8/4 | - Reduced inaccuracy by applying some localisation filters<br>- New mapping protocols |
| | | Correct (67.2%) 16cm | Passive Tags | 9m$^2$ | - System was not tested *on a* large scale with different zones.<br>- Lack of real time tracking for multiple objects. |
| **Jachimczyk et al [45], 2014 $^{S\&E}$** | Indoor positioning(3D localisation) | 7cm,11cm (based on 4 and 8 respectively readers) | TOA/RSS | 8/N/A | - More robustness and avoided obstacles<br>- Various configuration of active RFID Readers |
| | | 49cm, 50cm (based on 4 and 8 readers) | 3D passive tags-Hybrid | 46.17$^3$ | - Higher cost depends on how many RFID readers used *in the* configuration.<br>- High system complexity and more computational cost based on the scenarios |
| **Bolic et al [48]$^E$** | Indoor localisation (proximity detection) | 32 cm | Proximity | 2 /N/A | Inexpensive UHF RFID tags and they are maintenance free |
| | | 48 cm | Passive tags | 4m*2m | - Requirement of landmark tags for localisation application<br>- Relying on semi-passive tags (needs battery changes) |
| **Alsinglawi et al [24] $^E$** | Location estimation in Healthcare settings | 16.5 cm | Trilateration/RSSI | 3/1 | - Good accuracy levels with minimum tracking resources<br>- Cost-effective |
| | | | Passive Tags | 2.75m* 3.0 m | Uncertainty at blind spot area due to limited coverage |

## 4. CONCLUSIONS

This paper reviewed some of the most adaptable techniques and algorithms in RFID localisation. These techniques were grouped according to their approaches: distance estimation, sense analysis, and proximity. The paper highlighted the benefits and drawbacks of each of these localisation techniques for both passive and active RFID systems. Also, it discussed the current challenges

International Journal of Computer Networks & Communications (IJCNC) Vol.9, No.1, January 2017facing the implementation of indoor RFID localisation such as the Tags-Reader behaviour related issues. Moreover, it compared both active RFID and passive RFID systems with regards to accuracy, localisation type, coverage and deployment area. The advantages and disadvantages of each localisation approach were also pinpointed.

This study reveals the need for a new or enhanced algorithm that improves the passive RFID localisation method. Future works should aim to address the current challenges in RFID systems and the poor precision issue in a realistic environment where factors such as multipath, Human body interferences, and low-power pose a challenge for its effective deployment and operation. Furthermore, future works will look into implementing a cost-effective hybrid technology solution that could potentially improve the accuracy of locating and tracking of peoples and objects in smart environments.

International Journal of Computer Networks & Communications (IJCNC) Vol.9, No.1, January 2017

**AUTHORS**

**Belal Alsinglawi** is currently pursuing his Master Research Degree in Computing From School of Computing, Engineering and Mathematics at Western Sydney University. His main research interests include: RFID, Localisation, Smart homes, Internet of Things, Human Computer Interaction and Healthcare ICT.

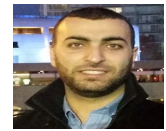

**Dr. Mahmoud Elkhodr** is with the School of Computing, Engineering and Mathematics at Western Sydney University (Western), Australia. He has been awarded the International Postgraduate Research Scholarship (IPRS) and Australian Postgraduate Award (APA) in 2012-2015. Mahmoud has been awarded the High Achieving Graduate Award in 2011 as well. His research interests include: Internet of Things, e-health, Human Computer- Interactions, Security and Privacy.

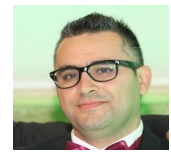

**Dr. Quang Vinh Nguyen** is a senior lecturer in Visual Analytics whose research interest is in Information Visualisation, Visual Analytics and Human Computer Interaction, particularly in genomic and biomedical and health data. During his academic career Nguyen has authored and co-authored over 80 refereed publications, including edited books, book chapters, journals and conference papers and has been successful in securing significant research grant funding.

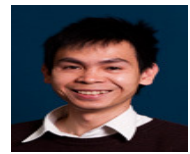

**Dr. Upul Gunawardana** obtained a BSc (Eng) (Hons 1) from the University of Moratuwa, Sri Lanka in 1990, MSEE and PhD from Missouri University of Science and Technology (formerly the University of Missouri-Rolla), USA in 1993 and 1999 respectively. Before joining Western Sydney University, he was with Philips Personal Communications as a member of their Technical Staff in 1998 and then with Motorola Personal Communication Systems Research Labs in New Jersey as a Staff Engineer from 1998 to January 2001. From 2001 to 2003, he was with Bell Laboratories of Lucent Technologies

**Professor Anthony Maeder** is Professor and Chair in Digital Health Systems at Flinders University. He founded the Telehealth Research and Innovation Laboratory (THRIL) at Western Sydney University in 2010. He is a Fellow of the Institution of Engineers Australia And a Fellow of the Australian Computer Society. He is currently a member of the Standards Australia IT-14 Health Informatics Committee. Professor Maeder was president of the Australasian Telehealth Society in the period 2010-2012, and joined the Governing Board of The International Society for Telemedicine and eHealth in 2013. In 2015 he was appointed as The Fulbright Distinguished Professor in Life Sciences and Agriculture at Kansas State University. 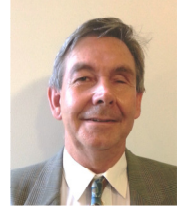

**Professor Simeon Simoff** is currently the Dean of the School of Computing, Engineering and Mathematics at the Western Sydney University. He is a founding Director and a Fellow of the Institute of Analytics Professionals of Australia. Currently he is an editor of the ACS Conferences in Research and Practice in Information Technology (CRPIT) series in Computer Science and ICT. Prior to that he was the associate editor (Australia) of the American Society of Civil Engineering (ASCE) Journal of Computing in Civil Engineering. Currently he serves on the ASCE Technical Committees on Data and Information Management, and on Intelligent Computing. 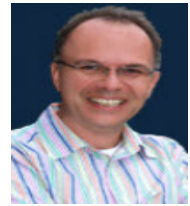